# Characterisation and Tribological Testing of Recycled Crushed Glass as an Alternative Rail Sand

Sadaf Maramizonouz[1], Sadegh Nadimi[1]*, William Skipper[2], Roger Lewis[2]
1 School of Engineering, Newcastle University, NE1 7RU, UK
2 Leonardo Centre for Tribology, Department of Mechanical Engineering, University of Sheffield, Sheffield, S1 3JD, UK

**ABSTRACT**

In the UK Network Rail Environmental Sustainability Strategy 2020-2050, minimal waste and the sustainable use of materials are highlighted as core priorities. The ambition is to reuse, repurpose or redeploy all resources. In low adhesion conditions, sand particles are used to enhance traction throughout the network. However, sand is in danger of becoming scarce as many applications demand it. In this study, an alternative adhesion enhancing particle system made of recycled crushed glass is examined in terms of density, size, shape distribution, mineralogy, mechanical properties, and bulk behaviour to better understand their characteristics in comparison with the typical Great British rail sand currently in use and reported in the literature. Their effects on tribological behaviour and surface damage are also investigated using the High-Pressure Torsion test in dry, wet, and leaf-contaminated conditions. Both particle characterisation and tribological testing show promising results. Recycled glass particles provide an acceptable level of traction with a similar level of rail damage as typical rail sand. It is suggested to perform full-scale laboratory and field tests to further confirm the suitability of this material.

* Corresponding author

Email: sadegh.nadimi-shahraki@newcastle.ac.uk

## 1. Introduction

The adhesion[1] or traction at the wheel-rail contact during train operation is of great importance. Loss of traction can lead to train delays, safety risks, and at worst accidents, resulting in a £345m cost to the British railway industry annually [1]. Undesirably low traction can be encountered when there are contamination layers on the top of the rail such as water or leaves. As one solution to increase traction to sufficient levels, rail sanding has been employed since the early years of the railway industry.

During the rail sanding process, sand particles are applied to the wheel-rail interface in a stream of compressed air targeted at the rail slightly ahead of the wheel as the train moves, the wheel passes over the sand particles resulting in their breakage and an increase in the wheel-rail traction. The efficiency of getting sand particles in right place is very low, resulting in around 80% sand wastage [2]. Sand resources are limited as this natural material is commonly used in many applications, such as computer microchips, construction, cosmetics, just to name a few. Therefore, sand may not be a practical option for rail sanding in the near future.

---

[1] In the railway industry "adhesion" or "adhesion coefficient" is defined as the amount of traction present when the wheel-rail contact enters partial slip. In this paper, the terms are used interchangeably.

Building on the Network Rail 30-year strategy for: delivering a sustainable railway, minimising the waste, and increasing the sustainable use of material [3], in this study, recycled crushed glass is proposed as an alternative to the standard sand used for rail sanding. The particles' density, shape and size, bulk behaviour, mechanical and mineralogical properties are characterised and compared to the typical rail sand. High Pressure Torsion (HPT) tests are executed to investigate the tribological performance of the crushed recycled crushed glass particles in dry, wet and leaf contaminated conditions. HPT results are compared to the data for the standard rail sand to present whether the crushed recycled crushed glass is a suitable substitute.

**2. Particle Characterisation**

*2.1 Preparation of recycled crushed glass particles*

Various types of glass bottles are used to produce the recycled crushed glass particles. The glass bottles are emptied and washed to remove the residual liquid and the labels and then crushed under compression. To produce the particles with desired sizes, ~100 grams of the crushed glass pieces are placed inside a laboratory disc mill (SIEBTECHNIK TEMA Machinery Scheibenschwingmühle TS 750) for 7 seconds. The resulting glass particles are sieved using three different mesh sizes to categorize them into three sieve cuts as follows: category 1 retained in a 2 mm mesh sieve (named Recycled Glass Large - RGL), category 2 retained in a 1.18 mm mesh sieve (RGM), and category 3 retained in a 600 μm mesh sieve (RGS). **Figure 1(a)** shows the size distribution of the particles from the three size categories compared to the standard Great British (GB) rail sand. It can be seen that the particle sizes are mostly inside the range currently proposed in GMRT 2461 which is from 0.71mm to 2.8mm [4]. The process is repeated until 2 kg of each sample is prepared.

2.2 *Density*

The density of particles is evaluated using the gas jar method following BS1377-2:1990 [5] using two samples of the recycled crushed glass particles each weighing ~400 grams. The average of the values of the density of the recycled crushed glass particles is 2512.18 kg/m$^3$ with a standard deviation of ±4.25 kg/m$^3$.

2.3 *Size and Shape Characterisation*

Particle shape characterisation is conducted by performing X-ray micro-Computed Tomography (μCT) scans. One sample from each particle size category, representative of the whole size distribution, is chosen and scanned utilising the μCT system (SkyScan 1176) located in the Preclinical in-Vivo Imaging Facility at Newcastle University Medical School, UK. The μCT system is operated with a source current of 357 μA and a voltage of 70 kV. The resulting μCT images are reconstructed to produce greyscale cross-sectional slices with a voxel side length/image resolution of 8.81 μm which resulted in a 3D image with 7444×7444×7117 voxels.

To make image analysis feasible, μCT images of each sample are resized with a factor of 0.25 resulting in a decrease in the size of the 3D matrix to 1861×1861×7117 voxels. This increases the computational efficiency, while making sure that the particles size and shape are conserved [6].

The images are analysed, and the particle shape descriptors are calculated from the 3D particle geometries using the SHAPE code by Angelidakis et al. [7], readers are referred to Angelidakis et al. [8] for more information on particle shape descriptors. Particle shape distributions are plotted on Zingg charts in terms of flatness and elongation and presented in **Figure 1(b)**, **(c)**, and **(d)** for category 1, 2, and 3, respectively.

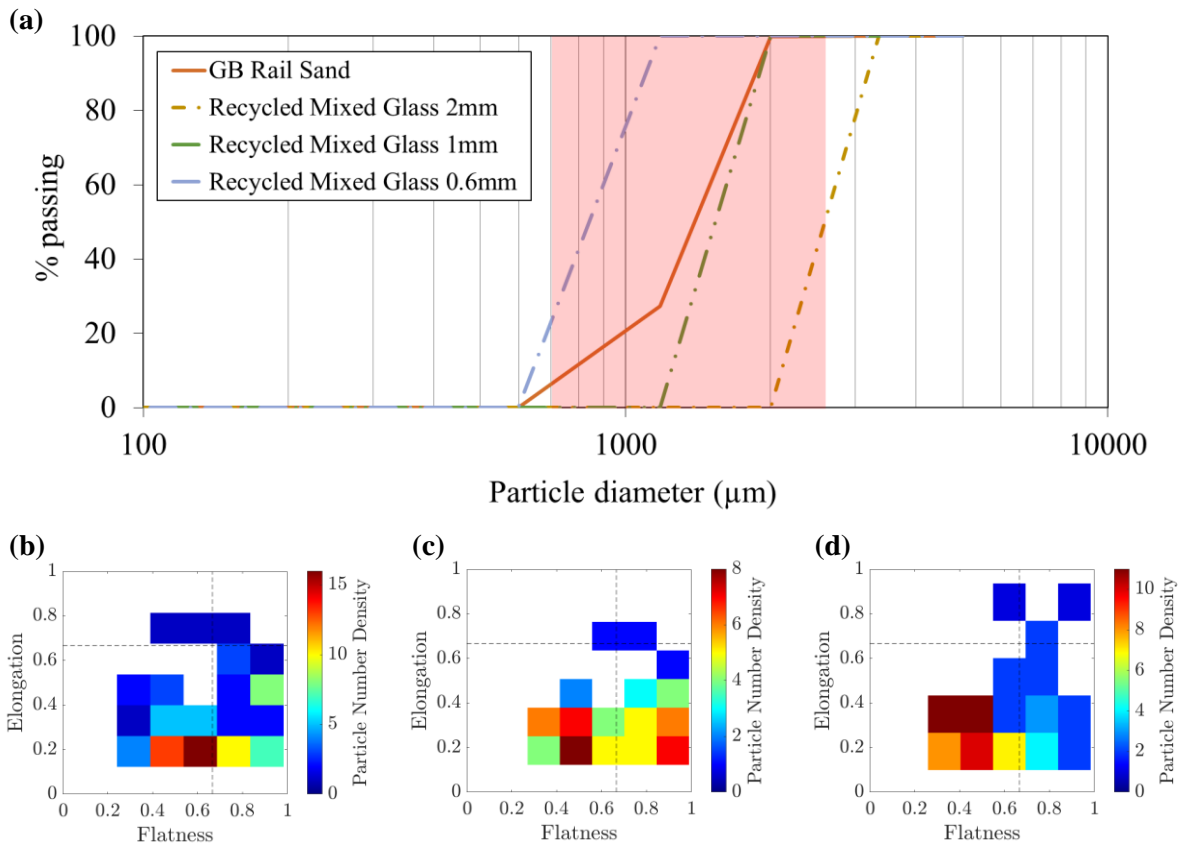

*Figure 1 (a) Particle size distribution of the three samples based on sieving. The red area shows the size range currently accepted by GMRT 2461[4]. Particle shape distributions of (b) sample 1 retained on 2 mm mesh sieve, (c) sample 2 retained on 1.18 mm mesh sieve, and (d) sample 3 retained on 600 μm mesh sieve based on flatness and elongation plotted on Zingg charts obtained from X-ray Computed Tomography.*

### 2.4 Bulk Characteristics

The Angle of Repose (AoR), the angle a pile of granular material produces relative to the horizontal plane which is a measure for the flowability of granular materials, is quantified to characterise the particles' bulk behaviour. Among the various methods of measuring AoR, none is defined as the standard. Here, the procedure proposed by the technical committee of the International Society for Soil Mechanics and Geotechnical Engineering (ISSMGE TC105) as a part of a round robin testing programme [9] is used to evaluate the AoR of the particles in categories 1, 2, and 3. A digital camera (SLR camera Canon (Tokyo, Japan) EOS 60D 18 MP CMOS) with EF-S 18-200-mm lens is utilised to capture the photos and the open-source software Fiji-ImageJ is used to measure the angles.

For each category, the tests are performed three times and the average value, and the standard deviation of the data are reported as follows: 37.38°±1° for category 1, 36.47°±1° for category 2, and 38.00°±1° for category 3.

### 2.5 Mechanical Properties

Nano-indentation tests are performed on the particles to measure their hardness and reduced modulus. Samples of the particles are mounted on a steel stub and tested utilising a nano-indentation instrument (NanoTest Vantage) and a diamond Berkovich indenter. The tip shape of the indenter is calibrated using a fused silica reference sample prior to testing. The experiments are carried out with a maximum load of 80 mN, loading time of 8 s, unloading time of 4 s, and maximum load hold of 10 s. The tests are repeated to obtain at least 10 reasonable indentations that are not adversely affected by the surface roughness. **Figure 2** presents the loading and unloading graphs for all the experiment instances. The particles' hardness (the ratio of maximum load to indentation area) and reduced modulus (the slope of unloading) are measured to be $7.28 \pm 0.12$ and $83.60 \pm 0.55$, respectively.

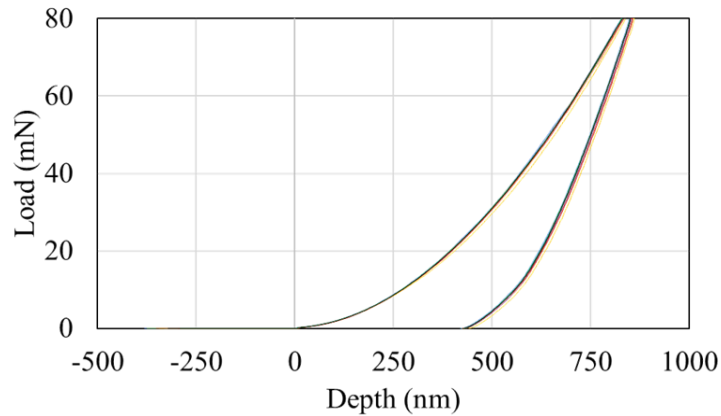

*Figure 2 Loading and unloading graphs during the nano-indentation tests for recycled crushed glass particles.*

2.6 *Mineralogical Properties*

For mineralogical characterisation and phase identification of the particles, powder X-ray diffraction (XRD) experiments are performed. For this purpose, a small sample of the particles is ground to a fine powder and transferred to the sample holder of the diffractometer (Bruker D2 Phaser with LynxEye detector using Cu Kα radiation). A preliminary scan (with 2theta between 5–100°) is run to check for low angle peaks prior to the main measurement scan. The diffractometer parameters are set to a divergence slit of 1.0 mm, with a 2theta range of 10–100°, step size of 0.033°, and 0.5 s step-1 and a Ni filter is used to reduce Kβ radiation. The detected peaks are compared to reference patterns for compounds/materials containing Si and O within the Crystallography Open Database. For the recycled crushed glass, although there is only one peak present within the pattern, it does match silicon oxide ($SiO_2$), (reference 96-900-5022). The X-ray diffraction pattern of the sample is presented in **Figure 3**.

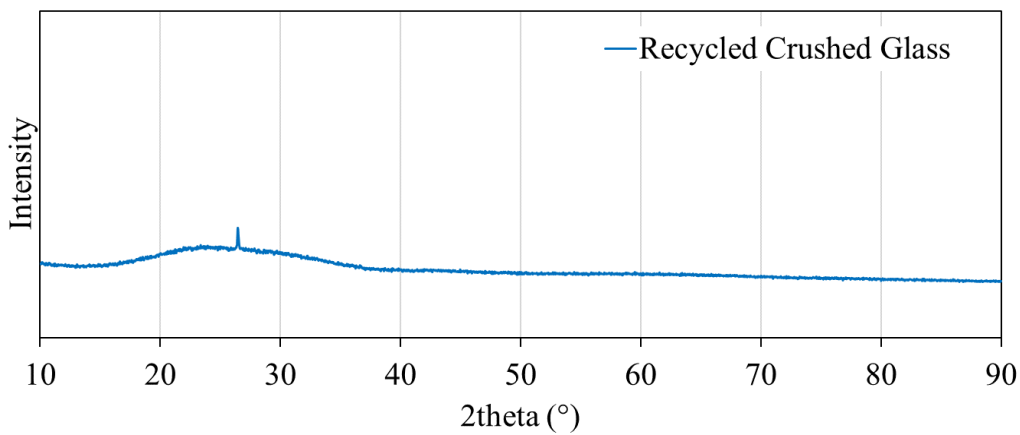

*Figure 3 Powder X-ray diffraction pattern for recycled crushed glass.*

## 3. Tribological Laboratory Testing

The set-up for the high-pressure torsion (HPT) experiments comprises of two flat specimens compressed together. Then, the contact is turned through a designated sweep angle by exerting a torque which varies for different contact conditions [10]. The HPT set-up located at the University of Sheffield is able to provide 400 kN and 1000 Nm of normal load and torque, replicating contact stresses with a maximum of 900 MPa, corresponding to a 60 kN load on one wheel [11].

One application of HPT tests is the investigation of the effects of sand on adhesion, as demonstrated by [1]. **Figure 4** presents a schematic of the HPT rig. The specimens made of wheel (1) and rail (2) steel are secured to their corresponding holders (3). First, there is no contact between the specimens, then, the two specimens are brought into contact and utilising the axial hydraulic actuator (5),a normal load is applied. A rotational hydraulic actuator (4) is employed to rotate the specimen faces against each other. Depending on the third body layer placed on the contact area of the specimens, the required amount of torque for turning the contact through the desired sweep angle is defined.

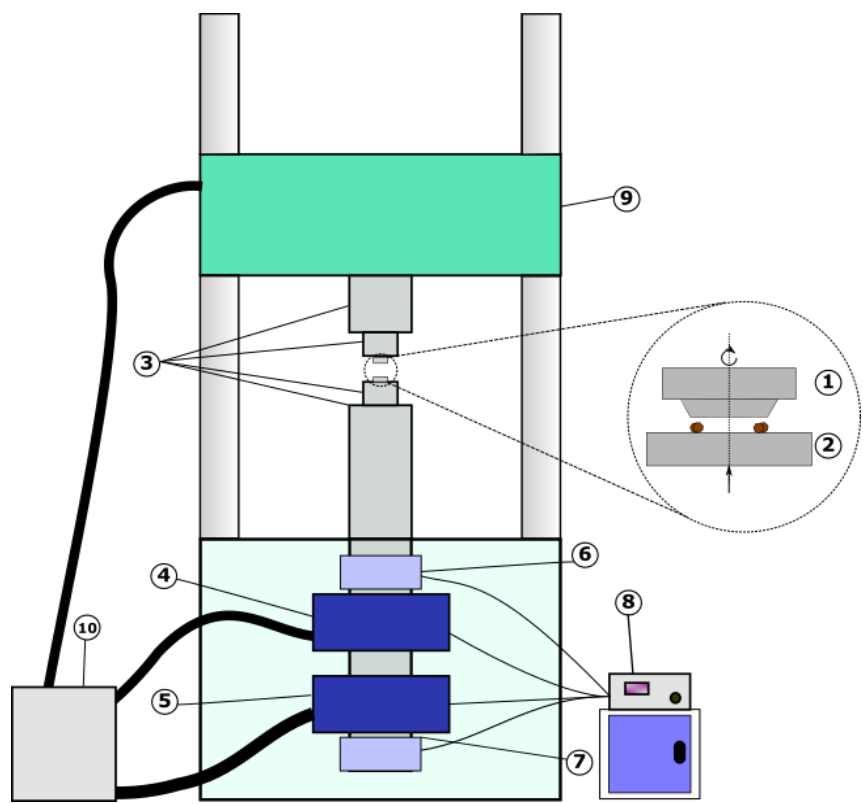

*Figure 4 Full Schematic of the High Pressure Torsion Rig (after [10])*

## 4. Concluding Remarks

**Figures 5(a)**, **(b)**, **(c)** show the HPT results in terms of coefficient of traction versus displacement for recycled glass (RG) in comparison with GB rail sand for three conditions of dry, wet and leaf contaminated contact. The surface roughness after the test was quantified and presented in **Figures 5(d)**, **(e)**, **(f)**. It can be seen that the level of traction and surface roughness by RG are comparable with GB for all conditions. Both materials have similar density, size, mechanical properties, and mineralogy (GB sand characteristics are reported in [1]). With regards to shape, GB particles are more compact, while RG is more elongated or bladed (see reference [8] for more information on particle shape classification). It is important to note that the AOR of RG is around 10 degrees higher than GB sands. This means that the flowability of RG is less than typical rail sand and may jam in the current sanding systems. Overall, it is suggested to perform full-scale laboratory and field tests to confirm the suitability of this material.

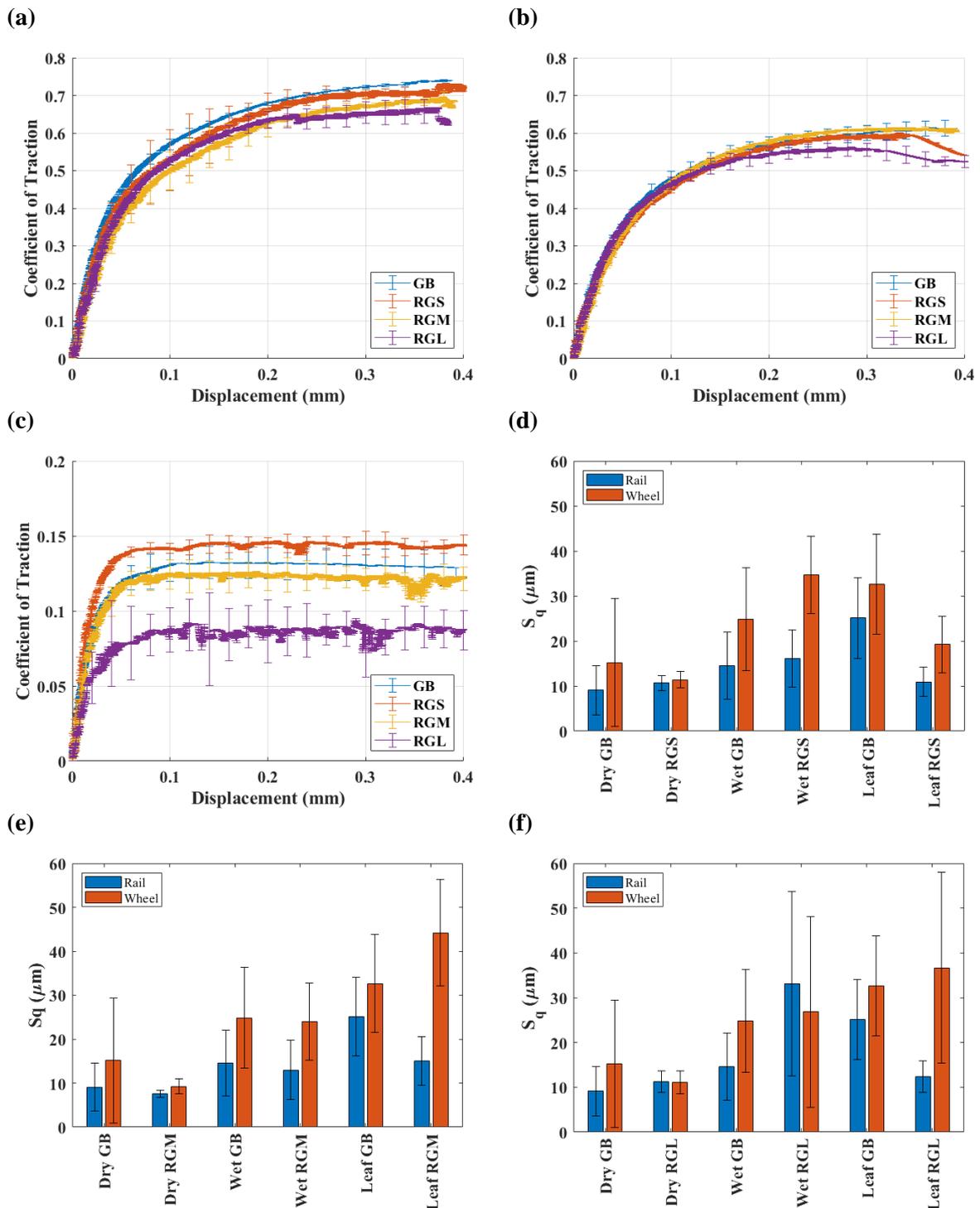

*Figure 5* High Pressure Torsion Results for (*a*) dry interface, (*b*) wet, and (*c*) leaf contaminated conditions; Surface roughness quantified using the Alicona InfiniteFocusSL 3D optical profilometer and reported in terms of root means square height of the surface roughness, (*d*) Recycled Glass Small (RGS), (*e*)Recycled Glass Medium (RGM), and (*f*) Recycled Glass Large (RGL).

**Acknowledgments**

This work is funded by the UK Engineering and Physical Sciences Research Council (EPSRC) grant No. EP/V053655/1 RAILSANDING - Modelling Particle Behaviour in the Wheel-Rail Interface.

**Conflict of Interest**

The authors declare that they have no known competing financial interests or personal relationships that could have appeared to influence the work reported in this paper.

**References**

[1] W. Skipper, S. Nadimi, A. Chalisey, and R. Lewis, "Particle Characterisation of Rail Sands for Understanding Tribological Behaviour," *Wear,* vol. 432, p. 202960, 2019.
[2] S. Lewis, S. Riley, D. Fletcher, and R. Lewis, "Optimisation of a Railway Sanding System for Optimal Grain Entrainment into the Wheel–Rail Contact," *Proceedings of the Institution of Mechanical Engineers, Part F: Journal of Rail and Rapid Transit,* vol. 232, no. 1, pp. 43-62, 2018.
[3] NetworkRail, "Network Rail Environmental Sustainability Strategy 2020-2050," 2020.
[4] *GMRT2461 Sanding Equipment (Issue 3)*, R. S. a. S. Board, 2018.
[5] *BS 1377-2:1990 Methods of test for soils for civil engineering purposes. Part 2: Classification tests*, B. S. Institution, 1990.
[6] V. Angelidakis, S. Nadimi, M. Garum, and A. Hassanpour, "Nano-Scale Characterisation of Particulate Iron Pyrite Morphology in Shale," *Particle & Particle Systems Characterization,* p. 2200120.
[7] V. Angelidakis, S. Nadimi, and S. Utili, "SHape Analyser for Particle Engineering (SHAPE): Seamless Characterisation and Simplification of Particle Morphology from Imaging Data," *Computer Physics Communications,* vol. 265, p. 107983, 2021.
[8] V. Angelidakis, S. Nadimi, and S. Utili, "Elongation, Flatness and Compactness Indices to Characterise Particle Form," *Powder Technology,* vol. 396, pp. 689-695, 2022.
[9] H. Saomoto and e. al, "Round robin test on angle of repose: DEM simulation results collected from 16 groups around the world," *Soils and Foundations,* vol. SANDF-S-22-00437, 2022.
[10] M. Evans, W. Skipper, L. Buckley-Johnstone, A. Meierhofer, K. Six, and R. Lewis, "The development of a high pressure torsion test methodology for simulating wheel/rail contacts," *Tribology International,* vol. 156, p. 106842, 2021.
[11] L. Zhou, H. Brunskill, R. Lewis, M. Marshall, and R. Dwyer-Joyce, "Dynamic characterisation of the wheel/rail contact using ultrasonic reflectometry," in *Proceedings of Railways 2014, The Second International Conference on Railway Technology: Research, Development and Maintenance*, 2014, pp. 8-11.